\documentclass[aps,prl,twocolumn,groupedaddress]{revtex4-1}
\usepackage{graphicx}

\begin{document}

\begin{flushright}
OU-HET-702
\end{flushright}


\title{Gauge-independence of gluon spin in the nucleon and its evolution}


\author{M.~Wakamatsu}
\affiliation{Department of Physics, Faculty of Science,
Osaka University, \\
Toyonaka, Osaka 560-0043, JAPAN}



\begin{abstract}
In recent papers, we have established the existence of gauge-invariant
decomposition of nucleon spin, each term of which can be related to
known high-energy deep-inelastic-scattering observables. A subtlety
remains, however, for the intrinsic spin part of gluons at the quantum level.
In fact, it was sometimes claimed that the evolution of gluon spin depends
on the gauge choice and its physical interpretation makes sense only in
the light-cone gauge. In the present paper, we will demonstrate explicitly
that the gluon spin operator appearing in our decomposition evolves
gauge-independently and that it properly reproduces the familiar evolution
equation for the 1st moments of polarized quark and gluon distributions
obtained with the Altarelli-Parisi method, which cannot directly be checked
by the standard operator expansion method.

\end{abstract}


\maketitle


In analyzing the decomposition of the nucleon spin, color gauge-invariance
is still the cause of intense
debate \cite{JM90}\nocite{Manohar90}\nocite{Manohar91}\nocite{Ji97PRL}
\nocite{HJL99}\nocite{BJ99}\nocite{SW00}\nocite{BLT04}\nocite{Chen08}
\nocite{Chen09}\nocite{Wakamatsu10}\nocite{Wakamatsu11}\nocite{Cho10}
\nocite{Wong10}\nocite{Leader11}-\cite{Hatta11}.
Although it has long been believed that the total angular momentum
of the gluon cannot be gauge-invariantly decomposed
into its intrinsic spin and orbital parts \cite{JM90},\cite{Ji97PRL},
this mistaken belief has recently been
questioned \cite{Chen08}-\cite{Wakamatsu11}.
In fact, we now have two different
gauge-invariant decompositions of nucleon spin into the spin and orbital
angular momenta of quarks and gluons. The one is the decomposition
proposed by Chen et al. \cite{Chen08},\cite{Chen09} and the other is the
one advocated by the present author \cite{Wakamatsu10},\cite{Wakamatsu11}.
A crucial advantage of the
latter decomposition is observability of each term of decomposition
by means of known high-energy deep-inelastic-scattering (DIS)
measurements \cite{Wakamatsu11}.

The circumstances for the quark part should already be clear from the
pioneering works by Ji \cite{Ji97PRL},\cite{Ji98JPG}.
First, the decomposition of the nucleon spin into the quark and gluon
total angular momentum, $J_q$ and $J_g$, is manifestly gauge-invariant,
and each term can in principle be extracted from
generalized-parton-distribution (GPD) analyses. Since the intrinsic
quark spin (or the longitudinal quark polarization)  $\Delta \Sigma$ is
clearly gauge-invariant and measurable,
the quark orbital angular momentum (OAM) defined as
$L_q \equiv J_q - \frac{1}{2} \,\Delta \Sigma$ should also be
gauge-invariant. An important fact here is that the quark OAM obtained
in that way corresponds to a nucleon matrix element of {\it dynamical}
quark OAM operator containing the full gauge-covariant
derivative \cite{Ji97PRL},\cite{Ji98JPG}, not the
{\it canonical} OAM or its nontrivial gauge-invariant extension as
advocated by Chen et al. \cite{Chen08},\cite{Chen09}.

Although slightly more delicate, the situation for the gluon part appears
quite analogous. It is widely believed that the gluon spin (or the longitudinal
gluon polarization) $\Delta g$ is a measurable quantity from the polarized
DIS experiments. Then, if one defines the gluon OAM by
$L_g \equiv J_g - \Delta g$, $L_g$ is clearly a measurable quantity.
A key question here is whether $\Delta g$ is really a gauge-invariant
quantity or not.
If it is indeed the case (as is vaguely anticipated), it is a logical
conclusion that $L_g$ is also gauge-invariant.
In this sense, a convincing check of the gauge-invariance
of gluon polarization in the nucleon is a fundamentally important
homework left in the nucleon spin physics.
The gauge-invariant gluon spin operator was first
proposed by Chen et al. \cite{Chen08},\cite{Chen09}, and it was later
confirmed by us \cite{Wakamatsu10},\cite{Wakamatsu11}.
Our investigation goes further. We could show in \cite{Wakamatsu11}
that the gluon OAM $L_g$
defined by $L_g \equiv J_g - \Delta g$, or more precisely the difference
between the 2nd moment of GPD $H_g (x,\xi,t) + E_g (x,\xi,t)$
and the 1st moment of the polarized gluon distribution $\Delta g(x)$,
coincides with the nucleon matrix element of the {\it dynamical} gluon
OAM containing the {\it potential angular momentum} term explained
in \cite{Wakamatsu10}, not the canonical OAM \cite{Leader11}
or its nontrivial gauge-invariant extension
advocated by Chen et al. \cite{Chen08},\cite{Chen09}

In spite of the above-mentioned nice correspondence between the
quark and gluon sectors, there still remains a subtlety in the gluon part.
It is no wonder that the gluon spin operator given in \cite{Wakamatsu11}
is gauge-invariant at least formally, or at the classical level.
It was also verified in \cite{Wakamatsu11} that
its nucleon matrix element reduces to the 1st moment of $\Delta g(x)$
in the LC gauge. If this is the case, a general thinking
based on the gauge pinciple dictates the following.
Since our gluon spin operator (although not necessarily local)
is gauge-invariant, its numerical value should be the same also in
any other gauges than the LC gauge.
Furthermore, its $Q^2$-evolution should also be
gauge-independent, as long as the regularization maintains gauge-invariance.

Unfortunately, life is not so simple. In fact, in an influential
paper \cite{HJL99}, Hoodbhoy, Ji, and Lu claimed that the gluon spin
evolves differently in two gauges, i.e. the LC gauge and the Feynman
gauge (a typical covariant gauge). According to them, the calculation in the
Feynman gauge does not reproduce the well-known Altarelli-Parisi
evolution equation for $\Delta \Sigma$ and
$\Delta g$ \cite{AP77}. Undoubtedly,
the difficulty is connected with the widely-known observation that
there is no gauge-invariant twist-2 local operator corresponding to
the gluon spin.
A state of disorder has been strengthened further by the recent claim by
Wong et al.~\cite{Wong10}. (See also a similar assertion by Cho et al. for
the evolution of quark and gluon momentum fractions \cite{Cho11}.)
They claim that the gluon spin in their decomposition evolves differently
from the standard Altarelli-Parisi equation. This sounds strange to us, because
their definition of gluon spin can be thought of as a gauge-fixed form
of our more general (or formal) gauge-invariant expression given in the
paper \cite{Wakamatsu11}. If gauge-invariance is maintained at every stage
of manipulation, the answer should be the same as that obtained
in the LC gauge.

Now, the purpose of the present paper is to resolve the above-explained
puzzle. To this end, we shall explicitly calculate the 1-loop anomalous
dimension of the quark and gluon spin operators appearing in our decomposition
within the Feynman gauge and show that the answer precisely coincides with
that obtained in the LC gauge.
 
Among the four terms in the decomposition \cite{Wakamatsu11},
we concentrate here on the quark and gluon spin operators given by
\begin{eqnarray}
 M^{+12}_{q-spin} &=& \bar{\psi} \,\gamma^+ \,\gamma_5 \,\psi, \\
 M^{+12}_{g-spin} &=& 2 \,\mbox{Tr} \left[\,F^{+1} \,A^2_{phys} \ - \ 
 F^{+2} \,A^1_{phys} \,\right] .
\end{eqnarray}
The gauge-invariance of $M^{+12}_{q-spin}$ is self-evident, while
that of $M^{+12}_{g-spin}$ is ensured by the covariant
transformation property of the {\it physical} part of the gluon field under
a gauge transformation, i.e. $A^\mu_{phys} (x) \rightarrow U(x) \,
A^\mu_{phys} (x) \,U^{-1} (x)$.

We start with writing down a little more explicit form of our gluon
spin operator : 
\begin{equation}
 M^{+12}_{g-spin} \ = \ V_A \ + \ V_B \ + \ V_C, \label{Gspin}
\end{equation}
with
\begin{eqnarray}
 V_A &=& (\partial^+ A^1_a) \,A^2_{a,phys} \ - \ 
 (\partial^+ A^2_a) \,A^1_{a,phys} , \label{VA} \\ 
 V_B &=& - \,\left[\,
 (\partial^1 A^+_a) \,A^2_{a,phys} \ - \ 
 (\partial^2 A^+_a) \,A^1_{a,phys} \,\right] , \label{VB} \\
 V_C &=& g \,f_{abc} \,A^+_b \,(\,
 A^1_c \,A^2_{a,phys} \ - \ A^2_c \,A^1_{a,phys} \,) . \label{VC}
\end{eqnarray}
In the LC gauge ($A^+_a = 0$), which falls into the category called the
physical gauge, only the vertex $V_A$ survives that simplifies the work
process significantly \cite{JTH96}.
Note however that, in general covariant gauge including the
Feynman gauge, the gluon spin operator is not given by $V_A$ alone, but
it is a sum of the three pieces $V_A, V_B$ and $V_C$.
We point out that $V_B$ and $V_C$
have basically the same form as the operators $O_1 = - \,\int d^3 x \,
\nabla A^{+}_a \times \vec{A}_a$ and $O_2 = - \,\int d^3 x \,f_{abc} \,
A^+_c \vec{A}_b \times \vec{A}_a$, which,
together with the naive gluon spin operator $S_g^+$ in the LC gauge,
were considered in the Feynman gauge calculation of the gluon spin
evolution equation \cite{HJL99}.
However, there is a very delicate difference between $V_A, V_B, V_C$
in Eqs.(\ref{VA})-(\ref{VC}) and $S^+_g, O_1, O_2$ considered in \cite{HJL99}.
One should notice that
one of the gluon fields in our operators $V_A, V_B, V_C$ is its
physical component. (Note that this is an important factor, which ensures
the gauge-invariance of our gluon spin operator.)

The question is then how to introduce this unique feature
of our gluon spin operator into the Feynman
rules for evaluating corresponding anomalous dimensions.
Remember first that the gluon propagator in general covariant gauge is
given by
\begin{eqnarray}
 D^{\mu \nu}_{ab} (k) &=& \frac{i \,\delta_{ab}}{k^2 + i \epsilon} \,
 \sum_{\lambda = 1}^4 \,\varepsilon^\mu (k, \lambda) \,
 \varepsilon^\nu (k,\lambda) \nonumber \\
 &=& \frac{i \,\delta_{ab}}{k^2 + i \varepsilon} \,
 \left(\,- \,g^{\mu \nu} \ + \ (1 - \xi) \,
 \frac{k^\mu k^\nu}{k^2 + i \varepsilon} \,\right) , \ \ \ 
\end{eqnarray}
with $\xi$ being an arbitrary gauge parameter. The Feynman gauge
corresponds to taking $\xi = 1$, while the Landau gauge to $\xi = 0$.
Since one of the gluon field appearing in our gluon spin operator is
its physical part, we must replace the gluon propagator by
\begin{equation}
 \frac{i \,\delta_{ab}}{k^2 + i \epsilon} \,
 \sum_{\lambda = 1}^2 \,\varepsilon^\mu (k, \lambda) \,
 \varepsilon^\nu (k,\lambda),
\end{equation}
when one of the endpoint of gluon propagator is obtained by the
contraction with the physical part of $A_\mu$ in our gluon spin
operator. Here, we need a sum of the product of gluon polarization
vector over {\it two physical polarization states} (not including the
scalar and longitudinal polarization).
The answer is well-known \cite{BD65},\cite{DKS03}.
Given below is its covariant form, which
holds in an arbitrary Lorentz frame : 
\begin{eqnarray}
 T^{\mu \nu} &\equiv&  \sum_{\lambda = 1}^2 \,
 \varepsilon^\mu (k, \lambda) \,
 \varepsilon^\nu (k,\lambda) \nonumber \\
 &=& - \,g^{\mu \nu} \ + \ \frac{k^\mu \,n^\nu + n^\mu \,k^\nu}{n \cdot k}
 \ - \ n^2 \,\frac{k^\mu \,k^\nu}{(n \cdot k)^2}, \ \ \ 
\end{eqnarray}
where $n$ being an arbitrary four-vector subject to the condition
$n \cdot \varepsilon = 0$ and $n \cdot k \neq 0$.
For practical calculation, it is convenient to take $n$ to be a
lightlike four-vector with $n^2 = 0$. In this case, the modified gluon
propagator reduces to
\begin{eqnarray}
 \tilde{D}^{\mu \nu}_{ab} (k) &\equiv& 
 \frac{i \, \delta_{ab}}{k^2 + i \epsilon} \,
 \sum_{\lambda = 1}^2 \,
 \varepsilon^\mu (k, \lambda) \,
 \varepsilon^\nu (k,\lambda) \nonumber \\
 &=& \frac{i \, \delta_{ab}}{k^2 + i \epsilon} \,\left(\,
 - \,g^{\mu \nu} \ + \ \frac{k^\mu \,n^\nu + n^\mu \,k^\nu}{n \cdot k}
 \,\right) , \ \ \ \label{TransPropagator}
\end{eqnarray}
which precisely coincides with the gluon propagator in the LC gauge.
This does not mean that we are working in the LC gauge from the very
beginning. In fact, if we did so, there would be no contributions to the
anomalous dimensions from the operators $V_B$ and $V_C$.
As pointed out before, it is crucial to use
the above propagator only when one of the endpoint of the gluon
propagator is obtained by the contraction with the physical part
of $A_\mu$ in our gluon spin operator.
In other places, we should use the standard gluon
propagator, which, for instance in the Feynman gauge, is given by
\begin{equation}
 D^{\mu \nu}_{ab} (k) \ = \ \frac{i \,\delta_{ab}}{k^2 + i \epsilon} \,
 \left(\,- \,g^{\mu \nu} \,\right).
\end{equation}

\begin{figure*}[htb] \centering
\begin{center}
 \includegraphics[width=10.0cm]{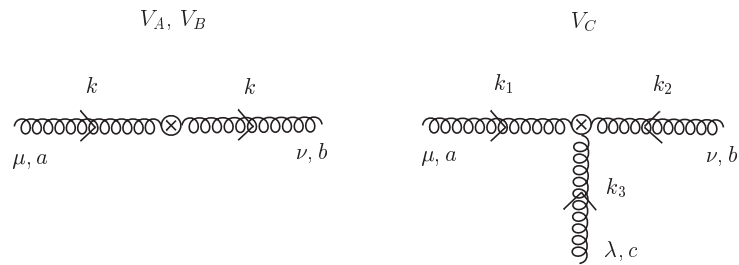}
\end{center}
\vspace*{-0.5cm}
\renewcommand{\baselinestretch}{1.20}
\caption{Momentum space vertices for the gluon spin operator.}
\label{Fig1:GspinVertex}
\vspace{6mm}
\end{figure*}

The momentum space vertex operators for the gluon spin, which
takes account of the above subtlety, can be expressed by the following
formulas supplemented with the diagrams shown
in Fig.\ref{Fig1:GspinVertex} : 
\begin{eqnarray}
 V_A &=& i \,k^+ \,(\, g^{\mu 1} \,g^{\nu 2} \, - \, g^{\mu 2} \,g^{\nu 1} \,)
 \,P^\nu_T \,\delta_{ab} \nonumber \\
 &-& \ (\mu \leftrightarrow \nu), \\
 V_B &=& - \,i \,g^{\mu +} \,(\, k^1 \,g^{\nu 2} \, - \, k^2 \,g^{\nu 1} \,)
 \,P^\nu_T \,\delta_{ab} \nonumber \\
 &-&  \ (\mu \leftrightarrow \nu), \\
 V_C &=& g \,f_{abc} \,g^{\lambda +} \,
 (\,g^{\mu 1} \,g^{\nu 2} \ - \ g^{\mu 2} \,g^{\nu 1} \,) \,
 (\,P^\mu_T \, + \, P^\nu_T \,) \nonumber \\
 &+& g \,f_{abc} \,g^{\mu +} \,
 (\,g^{\nu 1} \,g^{\lambda 2} \ - \ g^{\nu 2} \,g^{\lambda 1} \,) \,
 (\,P^\nu_T \, + \, P^\lambda_T \,) \nonumber \\
 &+& g \,f_{abc} \,g^{\nu +} \,
 (\,g^{\lambda 1} \,g^{\mu 2} \ - \ g^{\lambda 2} \,g^{\mu 1} \,) \,
 (\,P^\lambda_T \, + \, P^\mu_T \,) . \ \ \ \ \ 
\end{eqnarray}
Here, $P^\nu_T$ is a kind of projection operator, which reminds us of the
fact that we must use the gluon propagator $\tilde{D}^{\mu \nu}_{ab} (k)$
given by (\ref{TransPropagator}), whenever it contains the Lorentz
index $\nu$.

\begin{figure*}[htb] \centering
\begin{center}
 \includegraphics[width=12.0cm]{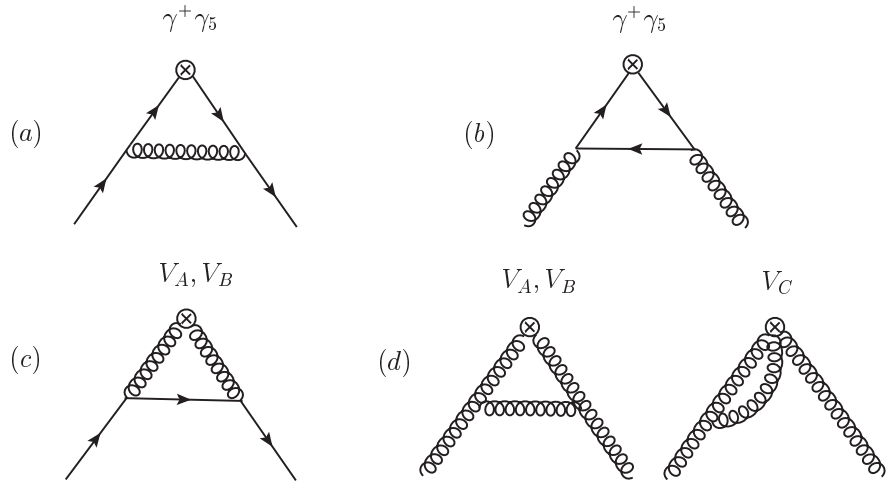}
\end{center}
\vspace*{-0.5cm}
\renewcommand{\baselinestretch}{1.20}
\caption{One-loop graphs contributing to the anomalous dimensions
of quark and gluon spin operators : (a) $\Delta \gamma^{(0)}_{qq}$,
(b) $\Delta \gamma^{(0)}_{qG}$, (c) $\Delta \gamma^{(0)}_{Gq}$, and 
(d) $\Delta \gamma^{(0)}_{GG}$. Graphs with external self-energies are
not shown in the figure. Graphs that are not symmeric with respect
to the vertical lines through the operator vertex have to be
counted twice.}
\label{Fig2:AnomalousDIM}
\end{figure*}

Keeping in mind somewhat nonstandard Feynman rule explained above,
we are now ready to calculate the 1-loop anomalous dimension of the
quark and gluon spin operators. The graphs $(a), (b), (c)$ and $(d)$,
shown in Fig.\ref{Fig2:AnomalousDIM}, respectively contribute to the
1-loop anomalous dimensions
$\Delta \gamma^{(0)}_{qq}, \Delta \gamma^{(0)}_{qG}, \Delta \gamma^{(0)}_{Gq}$,
and $\Delta \gamma^{(0)}_{GG}$. Graphs with external self-energies are
not shown in the figure. Graphs which are not symmetric with respect to
the vertical lines through the operator vertex have to be counted twice.

For the quark spin operator, we obtain just the expected answer : 
\begin{eqnarray}
 \Delta \gamma^{(0)}_{qq} &=& \frac{\alpha_S}{2 \,\pi} \cdot \frac{1}{2} \,C_F
 \ + \ \frac{\alpha_S}{2 \,\pi} \cdot \left(\,- \,\frac{1}{2} \,C_F \,\right)
 \ = \ 0  , \ \ \ \\
 \Delta \gamma^{(0)}_{qG} &=& \ \ \ 0,
\end{eqnarray}
with $\alpha_S = g^2 / (4 \,\pi)$ and $C_F = 4 / 3$.
Here, the 2nd term of $\Delta \gamma^{(0)}_{qq}$ comes from the
quark field-strength renormalization in the Feynman gauge.
For the gluon spin part, we have
\begin{eqnarray}
 \Delta \gamma^{(0)}_{Gq} &=& \frac{\alpha_S}{2 \,\pi} \cdot C_F \ + \ 
 \frac{\alpha_S}{2 \,\pi} \cdot \frac{1}{2} \,C_F \ = \ 
 \frac{\alpha_S}{2 \,\pi} \cdot \frac{3}{2} \,C_F , \ \ \ \  
\end{eqnarray}
where the 1st term is the contribution of the vertex $V_A$, while the
2nd is that of $V_B$. Finally, we find that
\begin{eqnarray}
 \Delta \gamma^{(0)}_{GG} &=& 
 \frac{\alpha_S}{2 \,\pi} \cdot \frac{11}{24} \,C_A
 \ + \ \frac{\alpha_S}{2 \,\pi} \cdot \left(\,- \,\frac{23}{24} \,C_A \right) 
 \nonumber \\
 &+& \frac{\alpha_S}{2 \,\pi} \cdot \frac{3}{2} \,C_A \ + \ 
 \frac{\alpha_S}{2 \,\pi} \cdot 
 \left(\,\frac{5}{6} \,C_A - \frac{1}{3} \,n_f \,\right)
 \ \ \ \nonumber \\
 &=& \frac{\alpha_S}{2 \,\pi} \cdot
 \left(\,\frac{11}{6} \,C_A - \frac{1}{3} \,n_f \right),
\end{eqnarray}
with $C_A = 3$ and $n_f$ being the number of quark flavors.
Here, the 1st, 2nd, and the 3rd terms are respectively the contributions
from the operators $V_A, V_B$ and $V_C$, whereas the 4th term comes
from the gluon field-strength renormalization in the Feynman gauge. 

One confirms that the final answer just coincides with the lowest-order
anomalous dimensions corresponding to the longitudinally polarized quark
and gluon distributions obtained with the Altarelli-Parisi method.
We emphasize that, in the standard operator expansion framework, a
direct confirmation of the above answer was done only in the LC gauge,
because there is no gauge-invariant local operator corresponding to the
gluon spin. As a consequence, the Altarelli-Parisi evolution equation for
the 1st moment of $\Delta q(x)$ and $\Delta g(x)$ has been justified only
on the basis of extrapolation or analytic continuation of higher
anomalous dimensions. Our calculation shows that, although
nonlocal, Eq.(\ref{Gspin})-(\ref{VC}) in fact gives the gauge-invariant
operator definition of the gluon spin, which reproduces the
Altarelli-Parisi evolution equation even in the Feynman gauge.
We can further show that the final answer for the
1-loop anomalous dimensions is exactly the same also in other covariant
gauges containing an arbitrary gauge parameter $\xi$. Here, the
dependence on the gauge parameter appears in the graph $(a)$ and
the left graph of $(d)$, but it is precisely cancelled by the $\xi$-dependent
terms arising from the quark and gluon field-strength renormalization.
In this way, the gauge-invariance of our gluon spin operator with
quantum-loop corrections is established.

In conclusion, we have calculated the 1-loop anomalous
dimensions of the gluon spin operator appearing in our gauge-invariant
decomposition of the nucleon spin \cite{Wakamatsu11}.
We find that the answer is
in fact independent of the choice of gauge. The answer obtained in the
Feynman gauge (or any covariant gauges) just reproduces the answer
obtained in the LC gauge, which is also the answer of the famous
Altarelli-Parisi method. The key factor leading to the correct answer
is the proper sum over the gluon polarization states appearing in the
Feynman rule. Note that such delicacy does not arise
in the Altarelli-Parisi method, in which only real splitting
processes containing physical gluons alone (not virtual gluons)
come into the calculation. At any rate, we have now confirmed that the
gluon spin operator appearing in our nucleon spin decomposition
provides us with the long-coveted operator definition of gluon spin,
which precisely reproduces the same evolution equation as obtained by
the Altarelli-Parisi method in an arbitrary gauge.
This means that, in combination with the previous two papers, we have
solved the long-standing puzzle concerning the
gauge-invariant decomposition of the nucleon spin.


\vspace{3mm}
\begin{acknowledgments}
This work is supported in part by a Grant-in-Aid for Scientific
Research for Ministry of Education, Culture, Sports, Science
and Technology, Japan, Grant No.~C-40135653.
\end{acknowledgments}


\end{document}